\documentclass{jfm}

\usepackage{subfigure}
\usepackage{graphicx}
\usepackage{natbib}
\usepackage{bm}

\ifCUPmtlplainloaded \else
  \checkfont{eurm10}
  \iffontfound
    \IfFileExists{upmath.sty}
      {\typeout{^^JFound AMS Euler Roman fonts on the system,
                   using the 'upmath' package.^^J}%
       \usepackage{upmath}}
      {\typeout{^^JFound AMS Euler Roman fonts on the system, but you
                   dont seem to have the}%
       \typeout{'upmath' package installed. JFM.cls can take advantage
                 of these fonts,^^Jif you use 'upmath' package.^^J}%
      }
  \else
  \fi
\fi

\ifCUPmtlplainloaded \else
  \checkfont{msam10}
  \iffontfound
    \IfFileExists{amssymb.sty}
      {\typeout{^^JFound AMS Symbol fonts on the system, using the
                'amssymb' package.^^J}%
       \usepackage{amssymb}%

      }{}
  \fi
\fi

\ifCUPmtlplainloaded \else
  \IfFileExists{amsbsy.sty}
    {\typeout{^^JFound the 'amsbsy' package on the system, using it.^^J}%
     \usepackage{amsbsy}}
    {\providecommand\boldsymbol[1]{\mbox{\boldmath $##1$}}}
\fi

\newsavebox{\astrutbox}
\sbox{\astrutbox}{\rule[-5pt]{0pt}{20pt}}

\title[Disentangle plume-induced anisotropy in the velocity field in buoyancy-driven turbulence]{Disentangle plume-induced anisotropy in the velocity field in buoyancy-driven turbulence}
\author[Zhou $\&$ Xia]{Quan ZHOU$^{1,2}$ \thanks{Email address for correspondence: qzhou@shu.edu.cn} and Ke-Qing XIA$^2$ \thanks{Email address for correspondence: kxia@phy.cuhk.edu.hk}}
\affiliation{$^1$Shanghai Key Laboratory of Mechanics in Energy and Environment Engineering, Shanghai Institute of Applied Mathematics and Mechanics, E-Institutes of Shanghai Universities, Shanghai University, Shanghai 200072, China\\
$^2$Department of Physics, The Chinese University of Hong Kong, Shatin, Hong Kong, China}
\pubyear{1996}
\volume{538}
\pagerange{119--126}
\date{?? and in revised form ??}

\begin{document}

\maketitle

\begin{abstract}

We present a method of disentangling the anisotropies produced by the cliff structures in turbulent velocity field and test it in the system of turbulent Rayleigh-B\'{e}nard (RB) convection. It is found that in the RB system the cliff structures in the velocity field are generated by thermal plumes.  These cliff structures induce asymmetry in the velocity increments, which leads us to consider the plus and minus velocity structure functions (VSF). The plus velocity increments exclude cliff structures, while the minus ones include them.  Our results show that the scaling exponents of the plus VSFs are in excellent agreement with those predicted for homogeneous and isotropic turbulence (HIT), whereas those of the minus VSFs exhibit significant deviations from HIT expectations in places where thermal plumes abound.  These results demonstrate that plus and minus VSFs can be used to quantitatively study the effect of cliff structures in the velocity field and to effectively disentangle the associated anisotropies caused by these structures.
\end{abstract}


\section{Introduction}

A paradigm for studying turbulent flows is the so-called homogeneous and isotropic turbulence (HIT) \cite[]{my1975}. Studying this idealized model allows one to focus on the essential physics of small-scale turbulence in the simplest possible case and use it as a first step to understand more complicated turbulence problems. However, in almost all flow systems existing in nature, anisotropy is always present and unavoidable. How to disentangle the effects of anisotropy in the experimentally or numerically measured physical quantities has been a major focus in turbulence research in recent years and several methods, such as the SO(3) group decomposition, have been put forward to separate the isotropic and anisotropic contributions in turbulent flows (Arad \emph{et al.} 1998, 1999; Grossmann, VON DER Heydt $\&$ Lohse 2001; Biferale \emph{et al.} 2002).

Buoyancy-driven turbulent flows occur widely in geophysical and astrophysical systems and in numerous engineering applications. In buoyancy-driven thermal turbulence, buoyant structures, such as thermal plumes, are predominant coherent structures that transport heat and drive the flow (Shang \emph{et al.} 2003; Xia, Sun $\&$ Zhou 2003). Earlier visualization experiments (Moses, Zocchi $\&$ Libchaber 1993; Xi, Lam $\&$ Xia 2004) have shown that these structures consist of a cap with sharp temperature gradient and a stem that is relatively diffusive and hence would generate cliff-ramp-like structures in temperature time series when passing a thermal probe \cite[]{belmonte1996pre, zhou2002prl}. It is well known that the so-called cliff-ramp structures would induce strong anisotropic effects. This has been widely studied in passive scalars \cite[]{warhaft2000arfm}, but the study on the effects of buoyancy-induced anisotropy is very limited. Here, we use turbulent Rayleigh-B\'{e}nard (RB) convection, a fluid layer heated from below and cooled on the top, as an example to study the anisotropic effects induced by buoyancy. In the past few decades, turbulent RB convection has become a model system for studying the phenomena and the generic physics associated with turbulent flows driven by buoyancy (Ahlers, Grossmann $\&$ Lohse 2009; Lohse $\&$ Xia 2010).

In the field of turbulence, the velocity structure functions (VSF), namely
\begin{equation}
S_p(r)=\langle|\delta_r v|^p\rangle,
\end{equation}
are usually used to characterize the turbulent kinetic energy cascades, and hence they are of prime importance and have been the central focus in the study of fluid turbulence (Sreenivasan $\&$ Antonia 1997; Ishihara, Gotoh $\&$ Kaneda 2009). Here, $\delta_r v = v(x+r)-v(x)$ is defined as the velocity increment over a separation $r$ and $\langle \cdots \rangle$ denotes an ensemble average. In particular,  the situation for thermally-driven turbulence is more complicated. \cite{B59} and \cite{O59}, denoted hereafter as BO59 for short, have long argued that within the so-called inertial range and above a certain buoyant scale, i.e. the Bolgiano length scale $\ell_B$, buoyant forces drive the cascade processes and scale as $S_p(r)\sim r^{3p/5}$. However, experimentally whether the BO59 scaling exists in turbulent RB system remains unsettled \cite[for a recent review, see][]{lx10arfm}. The maturing of spatial-velocity-field measuring techniques, e.g. the particle image velocimetry (PIV), has provided a great impetus for our understanding of such issues (Sun, Zhou $\&$ Xia 2006; Zhou, Sun $\&$ Xia 2008; Kunnen \emph{et al.} 2008; Zhou $\&$ Xia 2010). In particular, the work of \cite{sun2006prl} have shown that, at the center of the convection cell, the velocity field exhibits the same scaling behavior that one would find in HIT, whereas, near the cell sidewall, a ``new scaling" was found. While results at both places imply no BO59, the observed scaling behavior near sidewall remains unexplained by any existing theoretical models. We note that in a closed RB cell, the spatial distribution of the buoyancy-driven thermal plumes is highly inhomogeneous, they abound near the sidewall but are scarcely found at the center \cite[]{qiu2001prl, shang2003prl, shang2004pre, xi2004jfm}. These give us a clue that plumes may be responsible for the different scaling behavior observed at different places of the system and suggest that buoyancy plays an important role in the cascade process, but how this comes about is still missing. The objective of the present study is to address such a question, i.e., how buoyant forces influence the cascade properties in turbulent RB system?

\section{Experimental setup and parameters}

The convection cell has been described in detail elsewhere \cite[]{zx2010prl_lds, zhou2011jfm} and here we give only its main features. It is a vertical cylinder of height $H=50$ cm and inner diameter $D=50$ cm and thus of unity aspect ratio. The top and bottom plates are made of 1.5 cm thick pure copper with nickel-plated fluid-contact surface and the sidewall is made of a plexiglas tube of 5 mm in wall thickness. Deionized and degassed water was used as the convecting fluid. A square-shaped jacket made of flat plexiglas plates and filled with water is fitted to the outside of the sidewall, which greatly reduced the distortion effect to the PIV images caused by the curvature of the cylindrical sidewall.

\begin{figure}
\center
\resizebox{1\columnwidth}{!}{%
  \includegraphics{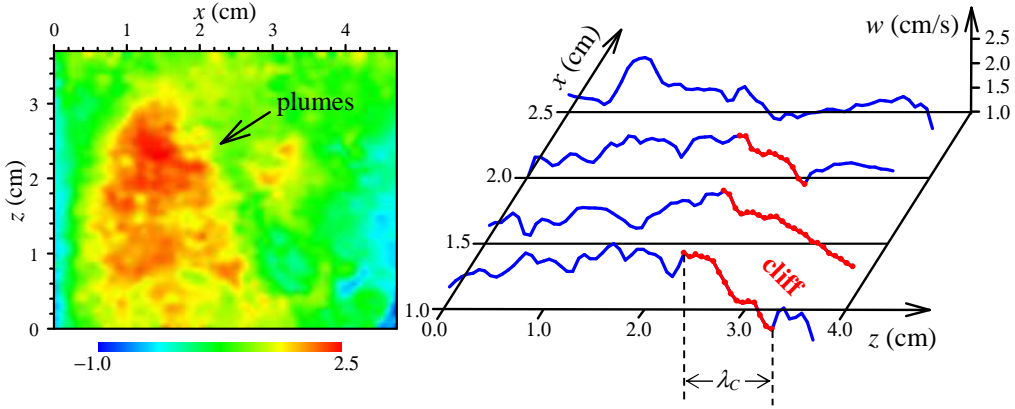}
} \caption{Left: A snapshot of the instantaneous vertical velocity field $w(x,z)$ near the cell sidewall for $Ra =4.0\times10^{10}$. Positive is defined for the upward motion and color coding is in cm/s. Right: Vertical slices of the vertical velocity component $w(x,z)$ through the left figure. The red curves mark the extracted cliff structures, corresponding to the regions of plume fronts.} \label{fig:fig1}
\end{figure}

Two series of measurements of the spatial velocity field were carried out using the PIV technique. In the first series, the measuring positions were fixed near the cell sidewall and the experiments covered the range $5.9\times10^9\lesssim Ra\lesssim1.1\times10^{11}$ of the Rayleigh number $Ra=\alpha g\Delta T H^{3}/\nu\kappa$, with $g$ being the gravitational acceleration, $\Delta T$ the temperature difference across the fluid layer, and $\alpha$, $\nu$ and $\kappa$ being, respectively, the thermal expansion coefficient, the kinematic viscosity, and the thermal diffusivity of water, whereas in the second series, the measurements were made from near the cell sidewall to the cell center at fixed Rayleigh number $Ra=4.0\times10^{10}$. For both series, the cell was tilted by a small angle of about 0.5$^{\circ}$ so that both series were made within the vertical plane of the large-scale circulation and at midheight of the cell. During the experiment the entire cell was wrapped by several layers of Styrofoam and the mean temperature of the convecting fluid was kept at $29^{\circ}$, corresponding to a Prandtl number $Pr=\nu/\kappa=5.5$. The details of the PIV measurement could be found elsewhere (Xia \emph{et al.} 2003; Sun, Xia $\&$ Tong 2005), here we give only its main features. Hollow glass spheres of 10 $\mu$m in diameter were chosen as seed particles and the thickness of the laser lightsheet  was $\sim0.5$ mm. The spatial resolution of the measured velocity field is 0.59 mm, which is much smaller than the lower end of the inertial range and hence is sufficient to reveal the scaling properties in the inertial range. In each measurement, the measuring region has an area of $4.7\times3.7$ cm$^2$ (see the left panel of figure \ref{fig:fig1}), corresponding to $79\times63$ velocity vectors, and the experiment lasted 3 hours in which a total of 25000 two-dimensional vector maps were acquired with a sampling rate $\sim2.3$ Hz. Since buoyant forces are exerted on the fluid in the vertical direction, we focus our attention mainly on the longitudinal vertical velocity increments $\delta_rw=w(x, z+r)-w(x, z)$, where $w(x, z)$ is the vertical velocity component obtained at position $(x, z)$. To acquire the accurate statistics, both temporal average and spatial average of moments of velocity increments within an area of $1.8\times3.7$ cm$^2$, i.e. $31\times63$ velocity vectors, were used when calculating VSFs, as the flow is approximately locally homogeneous in turbulent RB system \cite[]{sun2006prl,zhou2008jfm}.

\section{Results and discussions}

\subsection{Cliff structures in the vertical velocity field}

The left panel of figure 1 shows a typical snapshot of the vertical velocity field $w(x,z)$ near the cell sidewall, where $x$ is the horizontal distance from the wall. The arrow in the figure indicates the region with large vertical velocity, which is typically caused by a group of hot plumes passing by. The plumes are generally believed to be detached thermal boundary layers by buoyant forces. As introduced in $\S$ 1, thermal plumes usually generate cliff structures in the temperature field. Here, the snapshot of $w(x,z)$ and the associated slices through $w(x,z)$ in the vertical direction (the right panel of figure \ref{fig:fig1}) further show that in addition to temperature, plume fronts can also generate cliff structures in the vertical velocity field.  The formation of cliff structures in the velocity field may be understood as follows: Under the action of buoyant forces thermal plumes possess a higher speed in the vertical direction in comparison to the background fluids, which would deform the plumes such that they are compressed in the vertical direction and stretched in the horizontal directions. This deformation shortens the distance between the (relatively) high speed plume front and the low-speed downstream fluids, therefore resulting a steep velocity gradient, i.e. a cliff structure.

\begin{figure}
\center
\resizebox{1\columnwidth}{!}{%
  \includegraphics{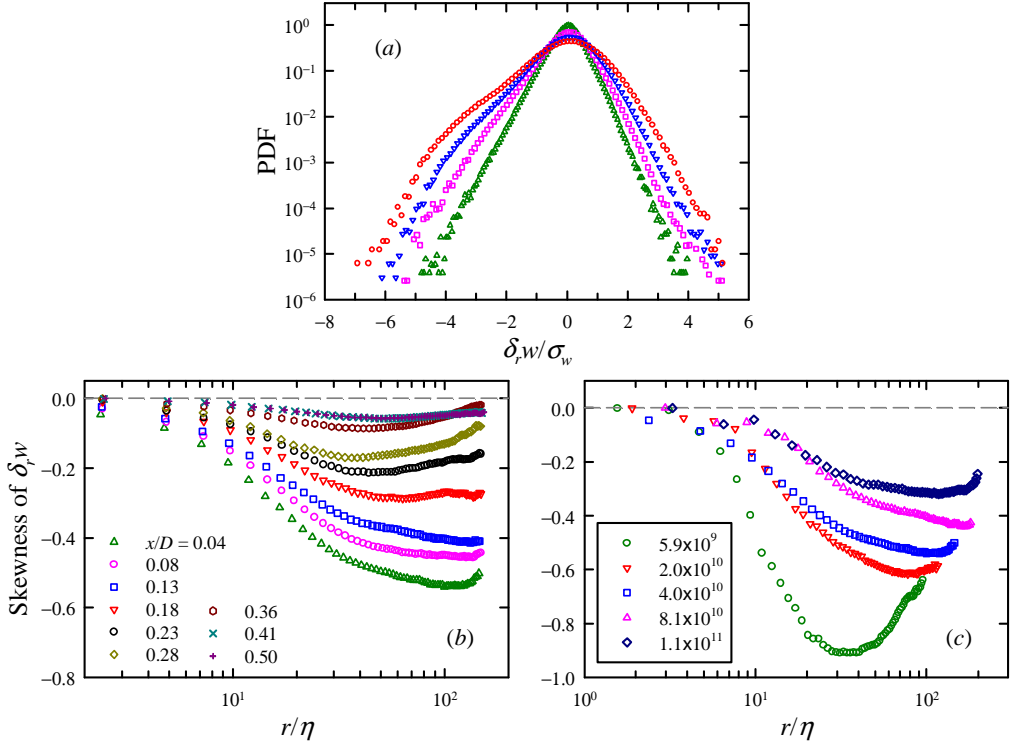}
} \caption{(\emph{a}) PDFs of the normalized increments of vertical velocity components at $x/D=0.04$. From the inner to the outer PDF, $r/\eta=16.6$ (dark-green up-triangles), 33.3 (pink squares), 57.1 (blue down-triangles), and 107 (red circles), all within the inertial range. (\emph{b}) Skewness of $\delta_rw$ as a function of $r/\eta$ for different measuring positions $x/D$ from near the cell sidewall to the cell center. The data were obtained at $Ra=4.0\times10^{10}$. (\emph{c}) Skewness of $\delta_rw$ as a function of $r/\eta$ for five different $Ra$ obtained near the cell sidewall ($x/D=0.04$).} \label{fig:fig2}
\end{figure}

One can expect that the persistence of such cliff structures would violate the local isotropy of turbulent flows. Isotropy is the central hypothesis for most theories and models of small-scale turbulence. Although a number of investigations in the atmosphere and in laboratory flows have shown that the skewness of velocity derivative is nontrivial \cite[]{sreenivasan97arfm}, implying local anisotropy at small scales, the presence of cliffs would enhance the degree of this small-scale anisotropy. This can be reflected by the probability density functions (PDF) of velocity increments over different scales. Figure \ref{fig:fig2}(\emph{a}) shows PDFs of $\delta_rw$ normalized by the standard deviation of $w$, $\sigma_w=\sqrt{\langle(w-\langle w\rangle)^2\rangle}$, over several different length scales within the inertial range. The data were obtained near the cell sidewall ($x/D=0.04$) at $Ra=4.0\times10^{10}$. The asymmetry of the distributions with long left tails is clear, especially for the (relatively) larger scales, which signifies the anisotropy at these scales. To quantify this anisotropy, we plot in figure \ref{fig:fig2}(\emph{b}) the skewness of $\delta_rw$ as a function of $r/\eta$ \footnote{The Kolmogorov length scale $\eta$ is estimated from $\eta=(\nu^3/\varepsilon)^{1/4}$, where $\varepsilon=\sigma_w^3/L$ is the energy dissipation rate per unit mass and $L$ is the largest length scale of the turbulence. It is found that $\eta$ changes from 0.27 mm near the sidewall to 0.26 mm at cell center for $Ra=4.0\times10^{10}$.} for different measuring positions ranging from $x/D=0.04$ (near the sidewall) to $x/D=0.50$ (at cell center). It is seen that all skewness are negative, but their magnitudes decrease continuously when moving away from the wall and appear to remain invariant for $x/D\gtrsim0.4$. If the nonvanishing skewness  observed here is indeed induced by cliff structures of thermal plumes, the behaviors of the skewness may then be understood from the inhomogeneous spatial distributions of thermal plumes in the convection cell, i.e., plumes abound near the sidewall but are scarce in the central region \cite[]{qiu2001prl, shang2003prl, shang2004pre, xi2004jfm}.

Figure \ref{fig:fig2}(\emph{c}) shows the skewness of $\delta_rw$ as a function of $r/\eta$ for five different $Ra$ obtained near the cell sidewall ($x/D=0.04$). Again, all skewnesses are found to be negative. Another noticeable feature is that the magnitudes of the skewness decrease with increasing $Ra$. This may be understood as the cliff structures are being smoothed out by the increased turbulent fluctuations. As the cliff structures of the velocity field are generated by buoyancy, this result suggests that temperature becomes more passive when the convective flow becomes more turbulent. Note that previous experimental study has shown that temperature may become progressively passive for $Ra>10^{10}$ \cite[]{belmonte1996pre, zhou2002prl}, which is qualitatively consistent with the picture obtained in the present study.

\subsection{Plus and minus velocity structure functions}

\begin{figure}
\center
\resizebox{1\columnwidth}{!}{%
  \includegraphics{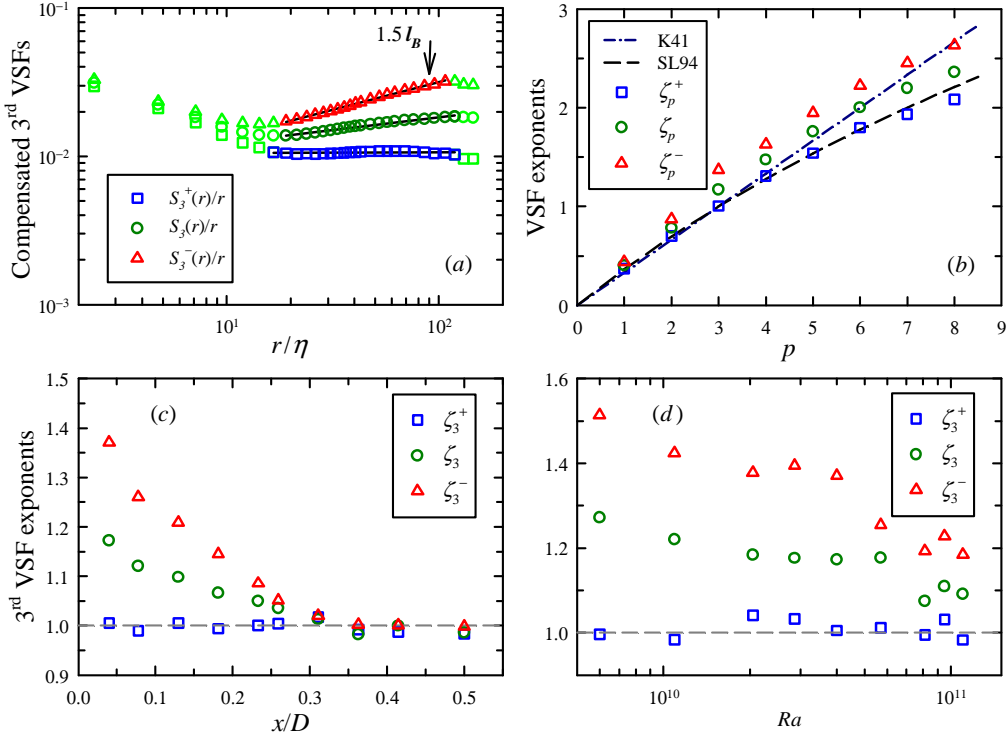}
} \caption{(\emph{a}) Compensated third-order VSFs $S_3^+(r)/r$, $S_3(r)/r$, and $S_3^-(r)/r$ measured near the cell sidewall. The arrow marks $r=1.5\ell_B$ for reference. (\emph{b}) Comparison of VSF exponents $\zeta_p^+$, $\zeta_p$, and $\zeta_p^-$ with various model predictions. The data were measured at $x/D=0.04$ for $Ra=4.0\times10^{10}$. (\emph{c}) The third-order VSF exponents $\zeta_3^+$, $\zeta_3$, and $\zeta_3^-$ as a function of the normalized distance from the cell sidewall $x/D$ for $Ra=4.0\times10^{10}$. (\emph{d}) $Ra$-dependence of $\zeta_3^+$, $\zeta_3$, and $\zeta_3^-$ obtained near the cell sidewall ($x/D=0.04$).} \label{fig:fig4}
\end{figure}

The fact that plumes can induce anisotropy suggests that the cascades of velocity field would possess different dynamics in different directions, i.e., the positive and negative velocity increments may have different scaling behaviors. To study this quantitatively, we examine the statistical properties of the plus and minus longitudinal VSFs \cite[]{sreenivasan1994prl, sreenivasan1996prl}, defined as
\begin{equation}
S_p^{\pm}(r)=\langle[(|\delta_rw|\pm\delta_rw)/2]^p \rangle.
\end{equation}
From this definition, it is clear that  cliff structures such as those shown in the right panel of figure  \ref{fig:fig1}) will be excluded from  $S_p^{+}(r)$ but included in  $S_p^{-}(r)$. This enables one to study separately the contributions of positive and negative velocity increments to the corresponding structure functions. A similar analysis has been performed previously in a turbulent channel flow and it was found that the plus VSFs are less affected by the presence of the wall \cite[]{onorate2001pre}. Figure \ref{fig:fig4}(\emph{a}) plots in log-log scale the compensated third-order VSFs $S_3^-(r)/r$ (triangles), $S_3(r)/r$ (circles), and $S_3^+(r)/r$ (squares) vs $r/\eta$ near the cell sidewall, which exhibit slightly different scaling ranges. The compensated plot shows a flat range for $S_3^+(r)/r$, i.e. $S_3^+(r)\sim r$, suggesting that $S_3^+(r)$ possesses the same scaling behavior as that for HIT, whereas both $S_3(r)$ and $S_3^-(r)$ exhibit much steeper scalings. To reveal this more clearly, we show in figure \ref{fig:fig4}(\emph{b}) the measured scaling exponents $\zeta_p^+$, $\zeta_p$, and $\zeta_p^-$ of the VSFs of orders $p=1$ to 8. When comparing $\zeta_p^+$ with the predictions of the hierarchy models of \cite{she1994prl} (SL94) for HIT, we find excellent agreement. These results suggest that the scaling behaviors of $S_p^+(r)$ are consistent with what one would expect for HIT. Moreover, because cliff structures contribute mainly to the negative velocity increments, $\zeta_p^-$ should exhibit some deviation from the K41-type scaling. This is indeed observed. Figure \ref{fig:fig4}(\emph{b}) shows that $\zeta_p^-$ and $\zeta_p$ are both much larger than the predictions of SL94.
Therefore, the plus and minus structures functions are effective means to study the effects of cliff structures in the velocity field and can be used to effectively disentangle the associated anisotropies caused by these structures. These cliff structures are produced by thermal plumes in the present case, but in general they can be produced by coherent structures in other types of flows.

Let's move on now to the location- and $Ra$-dependencies of the VSF scaling exponents. We focus mainly on the third-order because $\zeta_3=1$ is an exact result for HIT. Figure \ref{fig:fig4}(\emph{c}) shows the exponents $\zeta_3^+$, $\zeta_3$, and $\zeta_3^-$ as a function of the measuring position $x/D$. It is seen that both $\zeta_3$ and $\zeta_3^-$ are much larger than the K41-value of 1 near the cell sidewall and drop from the wall. For $x/D>0.3$ $\zeta_3$ and $\zeta_3^-$ are both essentially 1. On the other hand, apart from some data scatter, $\zeta_3^+$ assumes essentially the K41 value for all $x/D$. The decreases of $\zeta_3$ and $\zeta_3^-$ correspond to the reduced anisotropy associated to the cliff structures, i.e., the number of plumes decreases as the measuring position moves away from the cell sidewall, confirming quantitatively the results shown in figure \ref{fig:fig2}(\emph{b}). For $x/D>0.3$, i.e. the cell's central region, thermal plumes are scarce and hence it is not surprising to obtain approximately the K41 scaling for all three VSFs.

Figure \ref{fig:fig4}(\emph{d}) shows the measured scaling exponents versus $Ra$. Again, $\zeta_3^+$ is seen to remain nearly constant around the value of 1, but $\zeta_3$ and $\zeta_3^-$, despite certain scatter, show an overall decreasing trend with increasing $Ra$. Here, the decrease of $\zeta_3^-$ implies that the influence of buoyancy on the cascade processes becomes weaker when the flow becomes more turbulent, which could also be reflected by the behavior of the skewness of $\delta_rw$, whose magnitude is found to increase with decreasing $Ra$ as shown in figure \ref{fig:fig2}(\emph{c}).

\begin{figure}
\center
\resizebox{0.65\columnwidth}{!}{%
  \includegraphics{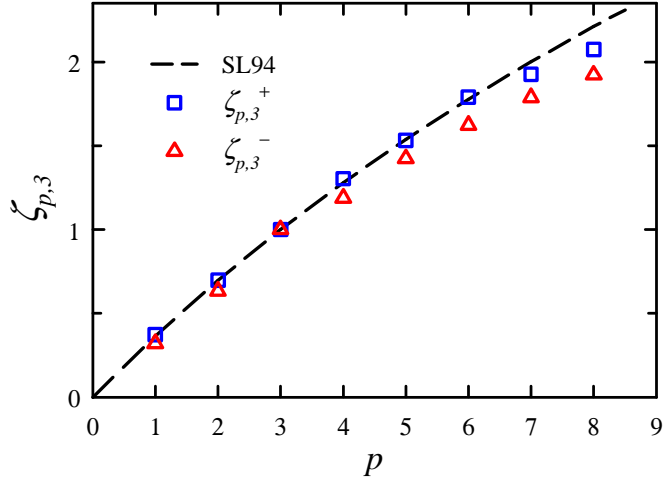}
} \caption{Comparison of ESS VSF exponents $\zeta_{p,3}^+$, $\zeta_{p,3}^-$, and the SL94 scaling exponents.} \label{fig:fig5}
\end{figure}

\begin{figure}
\center
\resizebox{0.65\columnwidth}{!}{%
  \includegraphics{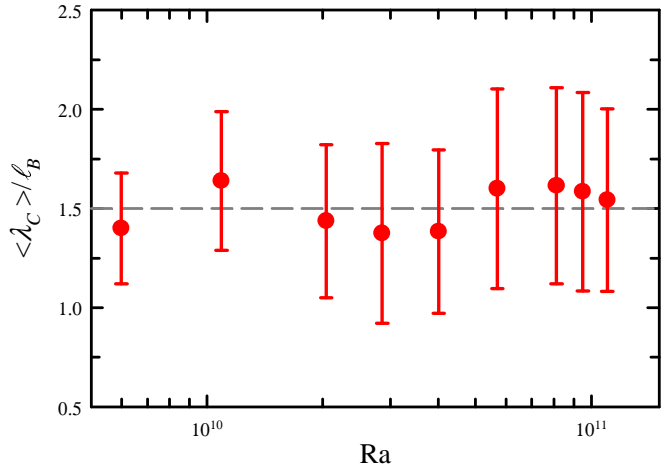}
} \caption{Mean cliffs' width $\langle\lambda_C\rangle$, normalized by the Bolgiano length scale $\ell_B$, as a function of $Ra$ obtained near the cell sidewall ($x/D=0.04$). The error bars mark the standard deviation of $\lambda_C$.}
\label{fig:fig6}
\end{figure}

It should be noted that near the sidewall the result $\zeta_3^-\gtrsim\zeta_3^+$, shown in figure \ref{fig:fig4}, does not imply that the positive velocity increments are more intermittent than the negative ones. To illustrate this, we examine the scaling behavior of VSFs via the extended self-similarity (ESS) method \cite[]{benzi1993pre}, i.e., $S_p(r)$ is plotted against $S_3(r)$, instead of $r$, in a log-log scale. Figure \ref{fig:fig5} shows the measured ESS (relative) scaling exponents $\zeta_{p,3}^+$ and $\zeta_{p,3}^-$ for the plus and minus VSFs, respectively. One sees that $\zeta_{p,3}^-$ are slightly smaller than $\zeta_{p,3}^+$. This result suggests that it is the minus velocity increments that possesses a higher degree of intermittency, which should be attributed to the persistence of cliff structures. A detailed analysis of such cliff structures would therefore be helpful for understanding the present results. To do this, we use a criterion to identify cliff structures in the vertical slices of $w(x,z)$ which is similar to those used for passive and active scalars \cite[]{tabeling2001prl, zhou2002prl}: a cliff is identified when $-\partial w/\partial z>\sigma_w/\ell_B$, where $\ell_B$ is based on the global quantities \cite[]{sun2006prl}. When such a cliff is found, we define its position $z_0$ as the point maximizing $|\partial w/\partial z|$, and its width $\lambda_C$ as the separation in space between the two extrema of $w$ surrounding $z_0$. Applying this procedure, over 50 000 cliffs were identified near the sidewall ($x/D=0.04$) for each $Ra$. Three examples of the extracted cliff structures are shown as red curves in the right panel of figure \ref{fig:fig1}. Figure \ref{fig:fig6} shows the mean cliffs' width $\langle\lambda_C\rangle$, normalized by $\ell_B$. One sees that $\langle\lambda_C\rangle=(1.5\pm0.2)\ell_B$ is nearly independent of $Ra$. We further note that the cliff structures mainly occur at the scales near the upper end of the VSF scaling range [see figure \ref{fig:fig4}(\emph{a})]. This could also be reflected qualitatively from figure \ref{fig:fig2}(\emph{a}), which shows that the distributions of $\delta_rw$ over large scales seem to be more asymmetric than those over relatively small scales, and quantitatively from figure \ref{fig:fig2}(\emph{b}), from which one sees that within the inertial range ($16\lesssim r/\eta\lesssim110$) the magnitude of the skewness of $\delta_rw$ increases with the scale $r$ for the near wall data ($x/D=0.04$, dark-green up-triangles). Cliff structures contain large velocity variations and would enhance the magnitude of velocity increments over the scales around their mean width $\langle\lambda_C\rangle$, i.e. the scale near the upper end of the VSF scaling range. This can also be seen from figure \ref{fig:fig4}(\emph{a}) that $S_p^+$ and $S_p^-$ differ the most around $1.5\ell_B$, signifying that scale is representative of the typical size of a cliff structure. The cliffs' influences to the scales near the lower end of the scaling range, however, are much weaker [see figure \ref{fig:fig4}(\emph{a})]. Therefore, under the influences of cliff structures, the value of $S_p^-(r)$ near the upper end of the VSF scaling range would increase, while those near the lower end do not change significant. As a result, the measured scaling exponents of $S_p^-(r)$ would increase. With decreasing buoyancy effects, the scaling exponents are also expected to drop, which is indeed observed in figure \ref{fig:fig4}(\emph{d}).

\section{Conclusion}

To summarize, we have demonstrated that as a manifestation of buoyancy effects thermal plumes can generate cliff structures in the vertical velocity field, which would in turn generate asymmetry in the velocity increments. We further show that such effects can be quantified by examining respectively the plus and minus velocity increments. For the plus increments, which largely exclude the cliff structures and hence removing the buoyancy effects, the scaling of their moments are the same as those expected of homogeneous and isotropic turbulence (HIT). For the minus increments, the scaling is found to be much steeper than that for HIT, which is caused by the presence of cliff structures. Such effects of buoyant forces are found to vanish gradually when moving away from the cell sidewall, owing to the inhomogeneous distributions of thermal plumes in a closed convection cell. It is also shown that, as a result of cliff structures being smoothed out by the increased turbulent fluctuations, the effect of buoyancy on the velocity field decreases with increasing $Ra$. Despite its simpleness, the analysis presented here provides a useful method to quantitatively study the effect of buoyancy and disentangle its contribution in the velocity field in buoyancy-driven turbulence. More generally, it may be used to quantify the effect of anisotropy in other complex phenomena when sharp fronts exist in the related random fields.

\begin{acknowledgments}
This work was supported in part by Natural Science Foundation of China (No. 11002085), ``Pu Jiang" project of Shanghai (No. 10PJ1404000), ``Cheng Guang" project of Shanghai (No. 09CG41), and E-Institutes of Shanghai Municipal Education Commission (Q.Z.) and by the Research Grants Council of Hong Kong SAR (Grant No. CUHK403806 and No. CUHK403807) (K.Q.X.).
\end{acknowledgments}


\begin{thebibliography}{32}
\expandafter\ifx\csname natexlab\endcsname\relax\def\natexlab#1{#1}\fi

\bibitem[Ahlers {\em et~al.\/}(2009)Ahlers, Grossmann \& Lohse]{agl09rmp}
{\sc Ahlers, G., Grossmann, S. \& Lohse, D.} 2009 Heat transfer and large scale
  dynamics in turbulent {Rayleigh}-{B}\'{e}nard convection. {\em Rev. Mod.
  Phys.\/} {\bf 81}, 503--537.

\bibitem[Arad {\em et~al.\/}(1999)Arad, Biferale, Mazzitelli \&
  Procaccia]{arad1999prl}
{\sc Arad, I., Biferale, L., Mazzitelli, I. \& Procaccia, I.} 1999
  Disentangling scaling properties in anisotropic and inhomogeneous turbulence.
  {\em Phys. Rev. Lett.\/} {\bf 82}, 5040--43.

\bibitem[Arad {\em et~al.\/}(1998)Arad, Dhruva, Kurien, L'vov, Procaccia \&
  Sreenivasan]{arad1998prl}
{\sc Arad, I., Dhruva, B., Kurien, S., L'vov, V.~S., Procaccia, I. \&
  Sreenivasan, K.~R.} 1998 Extraction of anisotropic contributions in turbulent
  flows. {\em Phys. Rev. Lett.\/} {\bf 81}, 5330--33.

\bibitem[Belmonte \& Libchaber(1996)]{belmonte1996pre}
{\sc Belmonte, A. \& Libchaber, A.} 1996 Thermal signature of plumes in
  turbulent convection: The skewness of the derivative. {\em Phys. Rev. E\/}
  {\bf 53}, 4893--98.

\bibitem[Benzi {\em et~al.\/}(1993)Benzi, Ciliberto, Tripiccione, Baudet,
  Massaioli \& Succi]{benzi1993pre}
{\sc Benzi, R., Ciliberto, S., Tripiccione, R., Baudet, C., Massaioli, F. \&
  Succi, S.} 1993 Extended self-similarity in turbulent flows. {\em Phys. Rev.
  E\/} {\bf 48}, R29--32.

\bibitem[Biferale {\em et~al.\/}(2002)Biferale, Lohse, Mazzitelli \&
  Toschi]{biferale2002jfm}
{\sc Biferale, L., Lohse, D., Mazzitelli, I. \& Toschi, F.} 2002 Probing
  structures in channel flow through {SO(3) and SO(2)} decomposition. {\em J.
  Fluid Mech.\/} {\bf 452}, 39--59.

\bibitem[Bolgiano(1959)]{B59}
{\sc Bolgiano, R.} 1959 Turbulent spectra in a stably stratified atmosphere.
  {\em J. Geophys. Res.\/} {\bf 64}, 2226--29.

\bibitem[Grossmann {\em et~al.\/}(2001)Grossmann, von~der Heydt \&
  Lohse]{gl2001jfm}
{\sc Grossmann, S., von~der Heydt, A. \& Lohse, D.} 2001 Scaling exponents in
  weakly anisotropic turbulence from the {Navier-S}tokes equation. {\em J.
  Fluid Mech.\/} {\bf 440}, 381--390.

\bibitem[Ishihara {\em et~al.\/}(2009)Ishihara, Gotoh \&
  Kaneda]{ishihara09arfm}
{\sc Ishihara, T., Gotoh, T. \& Kaneda, Y.} 2009 Study of high-{R}eynolds
  number isotropic turbulence by direct numerical simulation. {\em Annu. Rev.
  Fluid Mech.\/} {\bf 41}, 165--180.

\bibitem[Kunnen {\em et~al.\/}(2008)Kunnen, Clercx, Geurts, van Bokhoven,
  Akkermanns \& Verzicco]{kunnen2008pre}
{\sc Kunnen, R. P.~J., Clercx, H. J.~H., Geurts, B.~J., van Bokhoven, L. J.~A.,
  Akkermanns, R. A.~D. \& Verzicco, R.} 2008 Numerical and experimental
  investigation of structure-function scaling in turbulent
  {Rayleigh}-{B}\'{e}nard convection. {\em Phys. Rev. E\/} {\bf 77}, 016302.

\bibitem[Lohse \& Xia(2010)]{lx10arfm}
{\sc Lohse, D. \& Xia, K.-Q.} 2010 Small-scale properties of turbulent
  {Rayleigh}-{B}\'{e}nard convection. {\em Annu. Rev. Fluid Mech.\/} {\bf 42},
  335--64.

\bibitem[Moisy {\em et~al.\/}(2001)Moisy, Willaime, Andersen \&
  Tabeling]{tabeling2001prl}
{\sc Moisy, F., Willaime, H., Andersen, J.~S. \& Tabeling, P.} 2001 Passive
  scalar intermittency in low temperature helium flows. {\em Phys. Rev.
  Lett.\/} {\bf 86}, 4827--30.

\bibitem[Monin \& Yaglom(1975)]{my1975}
{\sc Monin, A.~S. \& Yaglom, A.~M.} 1975 {\em Statistical fluid mechanics\/}.
  vol. 2. MIT Press.

\bibitem[Moses {\em et~al.\/}(1993)Moses, Zocchi \& Libchaber]{moses1993jfm}
{\sc Moses, E., Zocchi, G. \& Libchaber, A.} 1993 An experimental study of
  laminar plumes. {\em J. Fluid Mech.\/} {\bf 251}, 581--601.

\bibitem[Obukhov(1959)]{O59}
{\sc Obukhov, A.~M.} 1959 On the influence of archimedean forces on the
  structure of the temperature filed in a turbulent flow. {\em Dokl. Akad.
  Nauk. SSSR\/} {\bf 125}, 1246--48.

\bibitem[Onorato \& Iuso(2001)]{onorate2001pre}
{\sc Onorato, M. \& Iuso, G.} 2001 Probability density function and ``plus" and
  ``minus" structure functions in a turbulent channel flow. {\em Phys. Rev.
  E\/} {\bf 63}, 025302(R).

\bibitem[Qiu \& Tong(2001)]{qiu2001prl}
{\sc Qiu, X.-L \& Tong, P.} 2001 Onset of coherent oscillations in turbulent
  {Rayleigh}-{B}\'{e}nard convection.

\bibitem[Shang {\em et~al.\/}(2003)Shang, Qiu, Tong \& Xia]{shang2003prl}
{\sc Shang, X.-D., Qiu, X.-L., Tong, P. \& Xia, K.-Q.} 2003 Measured local heat
  transport in turbulent {Rayleigh-B}\'{e}nard convection. {\em Phys. Rev.
  Lett.\/} {\bf 90}, 074501.

\bibitem[Shang {\em et~al.\/}(2004)Shang, Qiu, Tong \& Xia]{shang2004pre}
{\sc Shang, X.-D., Qiu, X.-L., Tong, P. \& Xia, K.-Q.} 2004 Measured local
  convective heat flux in turbulent {Rayleigh}-{B}\'{e}nard convection. {\em
  Phys. Rev. E\/} {\bf 70}, 026308.

\bibitem[She \& Leveque(1994)]{she1994prl}
{\sc She, Z.-S. \& Leveque, E.} 1994 Universal scaling laws in fully developed
  turbulence. {\em Phys. Rev. Lett.\/} {\bf 72}, 336--39.

\bibitem[Sreenivasan \& Antonia(1997)]{sreenivasan97arfm}
{\sc Sreenivasan, K.~R. \& Antonia, R.~A.} 1997 The phenomenology of
  small-scale turbulence. {\em Annu. Rev. Fluid Mech.\/} {\bf 29}, 435--472.

\bibitem[Sreenivasan {\em et~al.\/}(1996)Sreenivasan, Vainshtein, Bhiladvala,
  Gil, Chen \& Cao]{sreenivasan1996prl}
{\sc Sreenivasan, K.~R., Vainshtein, S.~I., Bhiladvala, R., Gil, I.~San, Chen,
  S. \& Cao, N.} 1996 Asymmetry of velocity increments in fully developed
  turbulence and the scaling of low-order moments. {\em Phys. Rev. Lett.\/}
  {\bf 77}, 1488--91.

\bibitem[Sun {\em et~al.\/}(2005)Sun, Xia \& Tong]{sun2005pre}
{\sc Sun, C., Xia, K.-Q. \& Tong, P.} 2005 Three-dimensional flow structures
  and dynamics of turbulent thermal convection in a cylindrical cell. {\em
  Phys. Rev. E\/} {\bf 72}, 026302.

\bibitem[Sun {\em et~al.\/}(2006)Sun, Zhou \& Xia]{sun2006prl}
{\sc Sun, C., Zhou, Q. \& Xia, K.-Q.} 2006 Cascades of velocity and temperature
  fluctuations in buoyancy-driven thermal turbulence. {\em Phys. Rev. Lett.\/}
  {\bf 97}, 144504.

\bibitem[Vainshtein \& Sreenivasan(1994)]{sreenivasan1994prl}
{\sc Vainshtein, S.~I. \& Sreenivasan, K.~R.} 1994 Kolmogorov's 4/5th law and
  intermittency in turbulence. {\em Phys. Rev. Lett.\/} {\bf 73}, 3085--88.

\bibitem[Warhaft(2000)]{warhaft2000arfm}
{\sc Warhaft, Z.} 2000 Passive scalars in turbulent flows. {\em Annu. Rev.
  Fluid Mech.\/} {\bf 32}, 203--240.

\bibitem[Xi {\em et~al.\/}(2004)Xi, Lam \& Xia]{xi2004jfm}
{\sc Xi, H.-D., Lam, S. \& Xia, K.-Q.} 2004 From laminar plumes to organized
  flows: {T}he onset of large-scale circulation in turbulent thermal
  convection. {\em J. Fluid Mech.\/} {\bf 503}, 47--56.

\bibitem[Xia {\em et~al.\/}(2003)Xia, Sun \& Zhou]{xia2003pre}
{\sc Xia, K.-Q., Sun, C. \& Zhou, S.-Q.} 2003 Particle image velocimetry
  measurement of the velocity field in turbulent thermal convection. {\em Phys.
  Rev. E\/} {\bf 68}, 066303.

\bibitem[Zhou {\em et~al.\/}(2011)Zhou, Li, Lu \& Liu]{zhou2011jfm}
{\sc Zhou, Q., Li, C.-M., Lu, Z.-M. \& Liu, Y.-L.} 2011 Experimental
  investigation of longitudinal space-time correlations of the velocity field
  in turbulent {Rayleigh-B}\'{e}nard convection. {\em J. Fluid Mech.\/} {\bf
  submitted}.

\bibitem[Zhou {\em et~al.\/}(2008)Zhou, Sun \& Xia]{zhou2008jfm}
{\sc Zhou, Q., Sun, C. \& Xia, K.-Q.} 2008 Experimental investigation of
  homogeneity, isotropy and circulation of the velocity field in
  buoyancy-driven turbulence. {\em J. Fluid Mech.\/} {\bf 598}, 361--372.

\bibitem[Zhou \& Xia(2010)]{zx2010prl_lds}
{\sc Zhou, Q. \& Xia, K.-Q.} 2010 Universality of local dissipation scales in
  buoyancy-driven turbulence. {\em Phys. Rev. Lett.\/} {\bf 104}, 124301.

\bibitem[Zhou \& Xia(2002)]{zhou2002prl}
{\sc Zhou, S.-Q. \& Xia, K.-Q.} 2002 Plume statistics in thermal turbulence:
  Mixing of an active scalar. {\em Phys. Rev. Lett.\/} {\bf 89}, 184502.

\end{thebibliography}

\end{document}